# Design and experimental demonstration of a high-directive emission with optical transformations

P-H. Tichit, S. N. Burokur, D. Germain, and A. de Lustrac[*]

[*]Affiliation: IEF, Univ. Paris-Sud, CNRS, UMR 8622, 91405 Orsay Cedex, France.

**With the explosion of wireless networks and automotive radar systems, there is an acute need for new materials and technologies that would not only minimize the size of these devices, but also enhance their performances. The technique of transformation optics—an innovative approach to produce artificial metamaterials that control electromagnetic waves as if space itself was transformed—provides unique opportunities to reach this goal. In this paper we design, fabricate and characterize a new class of metamaterial capable of transforming the source distribution and radiation pattern of an isotropic microwave emitter. Our findings have considerable implications for the development of new ultra-directive antennas with superior performances and compactness compared to conventional antennas operating in the same frequency range.**

The concept of transformation optics was first proposed by J. B. Pendry[1] and U. Leonhardt[2] in 2006. It provides the conceptual design of novel, and otherwise unattainable, electromagnetic and optical devices by controlling the paths of wave propagation. However, practical realization of these structures remains a challenge without the use of an extremely successful second concept, that of artificial metamaterials producing material parameters unobtainable in nature. The first example of this successful merging was the design and experimental characterization of an invisibility cloak in 2006[3]. Later other versions of cloaks have been proposed at microwave[4,5] and optical[6,7] frequencies. Invisibility-cloaking structures can serve as benchmark examples for the much broader ideas of transformation optics. In the last few years, the combination of transformation optics and metamaterials has led to staggering electromagnetic devices[8-19]. Proposals of new electromagnetic devices such as concentrators[9], wormholes[10], waveguide transitions and bends[11-15] and planar focusing antennas[16] have been theoretically submitted. Recently, experimental realizations and demonstrations on several transformation optics based devices have been conducted[17-19], but none in the domain of directive sources. Indeed, practical realization requires design and implementation of anisotropic metamaterials with high accuracy.

Although already known[20,21], the introduction of transformation optics in 2006 has allow to reappear the correspondence between coordinate transformation and materials parameters. In this way, the material can be viewed as a new geometry[1,2,22], and information about the coordinate transformation is given by material properties. Based on the reinterpretation of the form-invariance of Maxwell's equation against coordinate transformation, control of the electromagnetic field became possible at will by introducing a specific coordinate transformation that will map an initial space into an imagined one. Among the class of transformation found in literature, several possibilities are available for the design of electromagnetic structures. For example, continuous transformations introduced by Pendry to realise the first electromagnetic cloak[3] lead to anisotropic and inhomogeneous permeability and permittivity tensors but present the main advantage of being general. Contribution of these transformations was leveraged in many cases as cited just above. In parallel, Leonhardt proposed the concept of conformal mapping[2] where transformations respect Fermat's principle allowing design of devices with isotropic dielectric media[23-26]. The main drawback of such transformations is that mathematical requirement is often too complex for realization. Following this idea, quasi-conformal transformation[5,6,7,27-29] where slight deformation of the transformations can minimize the anisotropy of the material by approximation with an isotropic media was introduced. Other theoretical works mixed time and space transformations[29-32] and linked these transformations with cosmology and celestial mechanics[34-36]. At the same time, the concept of finite embedded transformation[11,12,37-40] was introduced adding a significant amount of flexibility and enabling steering or focusing electromagnetic waves. Finally, techniques of sources transformations[41-44] have offered new opportunities for the design of active devices with source distribution included in the transformed space. In this last approach, we design an ultra-directive emission by stretching an isotropic source into an extended coherent source.

The design step is very important, firstly because the values of theoretical electromagnetic parameters calculated by transformation optics are often too extreme to be realized; therefore a careful design must allow a reduction of these values. Also, in some cases permittivity and permeability tensors have non-diagonal terms that are difficult to implement. To facilitate the realization of structures, it is important to minimize or even cancel these terms. In all cases, for a real device the values of the electromagnetic parameters must be achievable with available metamaterials. Secondly, a practical realization necessitates a discretization of the theoretical material. This discretization must allow maintaining the performances of the theoretical structure at an acceptable level.

For instance, transformation optics has also allowed to transmute a singular isotropic profile into a regular but anisotropic material leading to a more practicable device. Dielectric singularities are points where the refractive index *n* reaches infinity or zero, where electromagnetic waves travel infinitely slow or infinitely fast. Such singularities cannot be made in practice for a broad spectral range, but one can transmute them into topological defects of anisotropic materials[18]. In reference Ma *et al.*[18] an omnidirectional retroreflector is implemented thanks to the transformation of a singularity in the index profile into a topological defect. Thus bounded values of the permittivity and permeability components allow the realization of the device. After the discretization step, tailored composite metal-dielectric metamaterials with optimized electromagnetic properties allow to approximate and implement these target distributions. Based on electric and magnetic resonances, these subwavelength structures can be appropriately engineered such that effective electromagnetic parameters can reach desired values.

Besides transformation optics approach, other interesting techniques have been proposed to achieve directive emissions. Enoch *et al.*[45] have shown how a simple stack of metallic grids can lead to ultra-refraction. Because the resulting metamaterial structure has an index of refraction, *n*, which is positive, but near zero, all of the rays emanating from a point source within such a slab of zero index material would refract, by Snell's Law, almost parallel to the normal of every radiating aperture. Another interesting metamaterial-based directive emission consisted in embedding a feed source between two parallel plate reflectors forming a resonant cavity antenna system[46].

In this article we present the practical implementation of a directive emission based on the transformation of an isotropic source at microwaves frequencies. Our aim is to show how a judiciously engineered metamaterial allows us to control the direction of emission of a source in order to collect all the energy in a small angular domain around the normal, with a good impedance matching between the radiating source and the material obtained by transformation optics. Following the theoretical procedure[1,2], constitutive electromagnetic parameter distributions are obtained for the material surrounding the radiating source. We describe the design of the anisotropic metamaterials used and the implementation of the proposed device. To experimentally demonstrate the directive emission, both the far field radiation pattern and the near field distribution are measured. Measurements agree quantitatively and qualitatively with theoretical simulations. The proposed device presents higher performances and compactness, compared to a parabolic reflector antenna with similar dimensions operating in the same frequency range. Our method though general is robust and can be easily extended to other frequency ranges and even at optical frequencies. The directive emitter finds important potential interests in communication systems for applications to high-rate data transmission, automotive radar, broadband point-to-point communications, and millimeter wave imaging.

**Design and Simulation**

In a topological approach, transformation optics consists in generating suitable metrics where light follows the geodesics. Here is a brief summary of the theoretical underlying physics of the transformation involved in this present work and the application to our antenna concept[44]: the imagined space of our proposed antenna is obtained by transforming a flat isotropic cylindrical half-space with zero Riemann curvature tensor described in polar coordinates {r, θ} into a flat space in squeezed Cartesian coordinates. x', y' and z' are taken to be the coordinates in the virtual transformed rectangular space and x, y, z are those in the initial real cylindrical space, as illustrated by the schematic principle in figure 1.

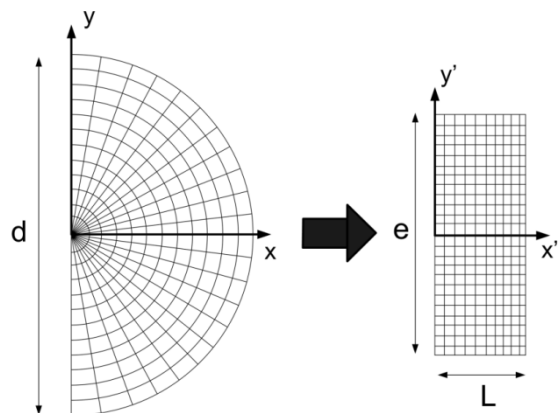

**Figure-1:** *Representation of the transformation of initial space into desired space. Each radius line of the cylindrical space is transformed into a horizontal line of the right rectangular space.*

We assume free space in the cylinder, with isotropic permeability and permittivity tensors $\varepsilon_0$ and $\mu_0$. Theory has shown that the coordinate transformation can be implemented by a material obeying the tensors

$$\theta^{i'j'} = g^{i'j'}\left|\det\left(g^{i'j'}\right)\right|^{-\frac{1}{2}}\theta \qquad (1)$$

where $\theta$ represents the permittivity or permeability tensor and $g$ the metric tensor of our designed space. To be implemented the material must be able to produce the following simple dielectric tensors:

$$\varepsilon^{ij} = \mu^{ij} = diag\left(\varepsilon_{xx}(x'), \frac{1}{\varepsilon_{xx}(x')}, \alpha\varepsilon_{xx}(x')\right) \qquad (2)$$

where $\varepsilon_{xx}(x') = \frac{\pi x'}{e}$ and $\alpha = \frac{d^2}{4L^2}$, with d representing the diameter of the initial cylindrical space and e and L, respectively the width and length of the rectangular target space. The appropriate choice of our transformation thus assures an absence of non-diagonal components, giving rise to a practical implementation using metamaterials. However this realization needs further simplifications of the material electromagnetic parameters. Firstly, the dimensions of the semi-cylindrical space must be set so that $\alpha = 4$ in order to obtain achievable values for the electromagnetic parameters. Additional simplification arises from the choice of the polarization of the emitted wave. Here we consider a polarized electromagnetic wave with an electric field pointing in the z-direction, which allows modifying the dispersion equation in order to simplify the electromagnetic parameters without changing Maxwell's equations and propagation in the structure. To obtain for these electromagnetic parameters values compatible with the manufacturing technology we use the same method as in reference 3. We multiply by $\mu_{xx}$ the dispersion equation and then our metamaterial is now simply described by:

$$\mu_{xx} = 1 \qquad \mu_{yy} = \frac{1}{\left(\varepsilon_{xx}\right)^2} \qquad \varepsilon_{zz} = 4\left(\varepsilon_{xx}\right)^2 \qquad \text{with e = 15 cm} \qquad (3)$$

The penalty of the above reduction is an imperfect impedance matching at the outer boundary of our metamaterial that we can evaluate as $Z = \sqrt{\frac{\mu_{yy}}{\varepsilon_{zz}}}(x=L) = \frac{9}{2\pi^2}$ with L=5cm and d=15cm. Thus the transmission at the outer boundary is calculated classically with $T = \frac{4Z}{(1+Z)^2} = 0.85$ which assures a high level of radiated electromagnetic field. Further simplification consists in discretizing the desired variation of the parameters $\mu_{yy}$ and $\varepsilon_{zz}$ to secure a practical realization producing experimental performances close to theory.

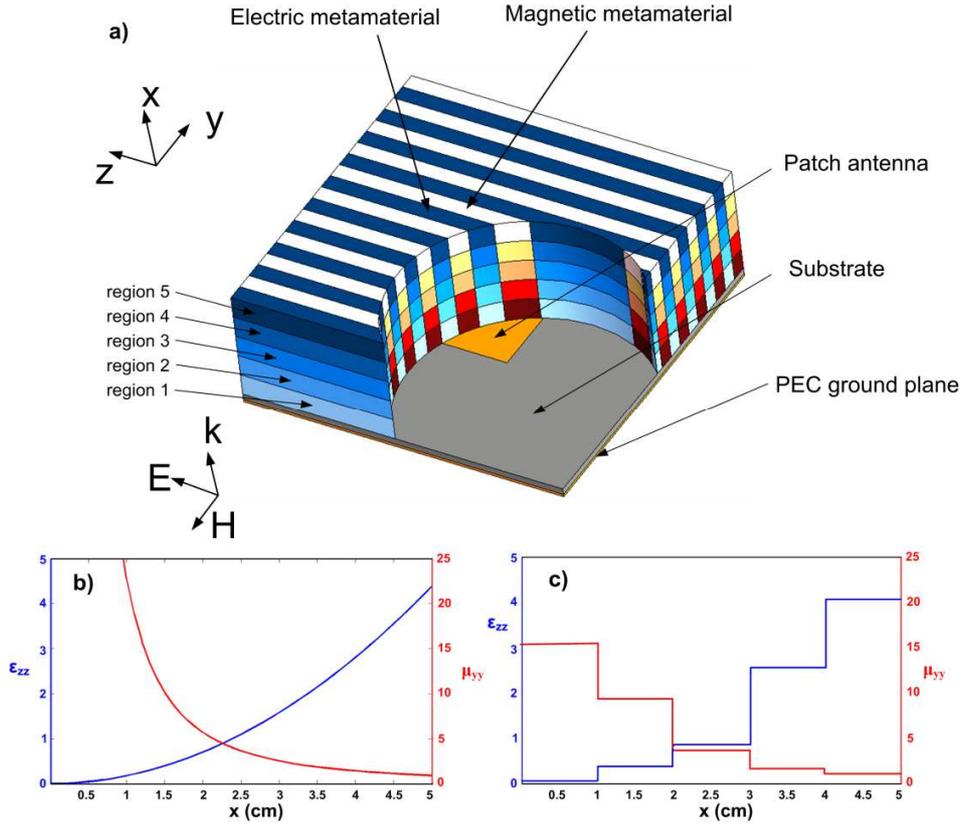

**Figure-2:** *(a) Schematic structure of the proposed antenna with a cylindrical cut to show the internal structure of the material. The emitting source is a microstrip patch antenna on a dielectric substrate. The metamaterial is composed of alternating permittivity and permeability vertical layers. Each layer is made of 5 different material regions (pale colour near the patch to dark colour in the x-direction). (b) Theoretical material parameters given by relation (3). (c) Discrete values of material parameters used in experimental realization.*

Fig. 2a shows the schematic structure of the directive emission antenna. A microstrip patch antenna on a dielectric substrate constitutes the radiating source. A surrounding material made of alternating electric metamaterial and magnetic metamaterial layers transforms the isotropic emission of the patch antenna into a directive one. The material is composed of five different regions where permittivity and permeability vary according to the profile of Fig. 2c. The corresponding reduced magnetic and electric properties of the metamaterial obtained from transformation optics are presented in Fig. 2b and 2c. The distribution of the theoretical material parameters satisfying relation (3) is shown in Fig. 2b. The distribution in Fig. 2c presents the discrete values corresponding to the five regions of the metamaterial used for the experimental validation. To implement the material specifications in Eq. (3) using metamaterials, we must choose the overall dimensions, design the appropriate unit cells, and specify their layout. For our implementation, the metamaterial unit cell is not periodic. It is also advantageous to optimize the three design elements all at once since common parameters are shared. Eq. (3) shows that the desired ultra-directive emission will have constant $\mu_{xx}$, with $\varepsilon_{zz}$ and $\mu_{yy}$ varying longitudinally throughout the structure. The axial permittivity $\varepsilon_{zz}$ and permeability $\mu_{yy}$ show respectively values ranging from 0.12 to 4.15 and from 1.58 to 15.3.

As shown by the schematic structure of the antenna in Fig. 2a, a square copper patch is printed on a 0.787 mm thick low-loss dielectric substrate (Rogers RT/Duroid 5870™ with 17.5 µm copper cladding, $\varepsilon$ = 2.33 and tan$\delta$ = 0.0012) and used as feeding source. The metamaterial covers completely the patch feeding source to capture the emanating isotropic radiation and transform it into a directive one. The metamaterial is a discrete structure composed of alternating layers with anisotropic permeability and permittivity. Fig. 3 shows a photography of our fabricated antenna device. We built the bulk metamaterial from 56 layers of dielectric boards on which subwavelength resonant structures are printed. 28 layers contain artificial magnetic resonators and 28 electric ones. Each layer is made of 5 regions of metamaterials corresponding to the discretized values of Fig. 2c. The layers are mounted 2 by 2 with a constant air spacing of 2.2 mm between each, in order to best represent the permeability and permittivity characteristics in the different regions. Overall dimensions of our antenna are 15 cm x 15 cm x 5 cm.

The details of the metamaterial cells are illustrated in Fig. 3. The left and right inserts show the designs of the resonators used in the magnetic (right) and electric (left) metamaterial layers. The layers are divided in 5 regions in the direction of wave propagation. Each region is composed of three rows of resonators with identical geometry and dimensions. Different resonators are used for electric and magnetic layers. Their schematic drawings are depicted at the bottom of Fig. 3.

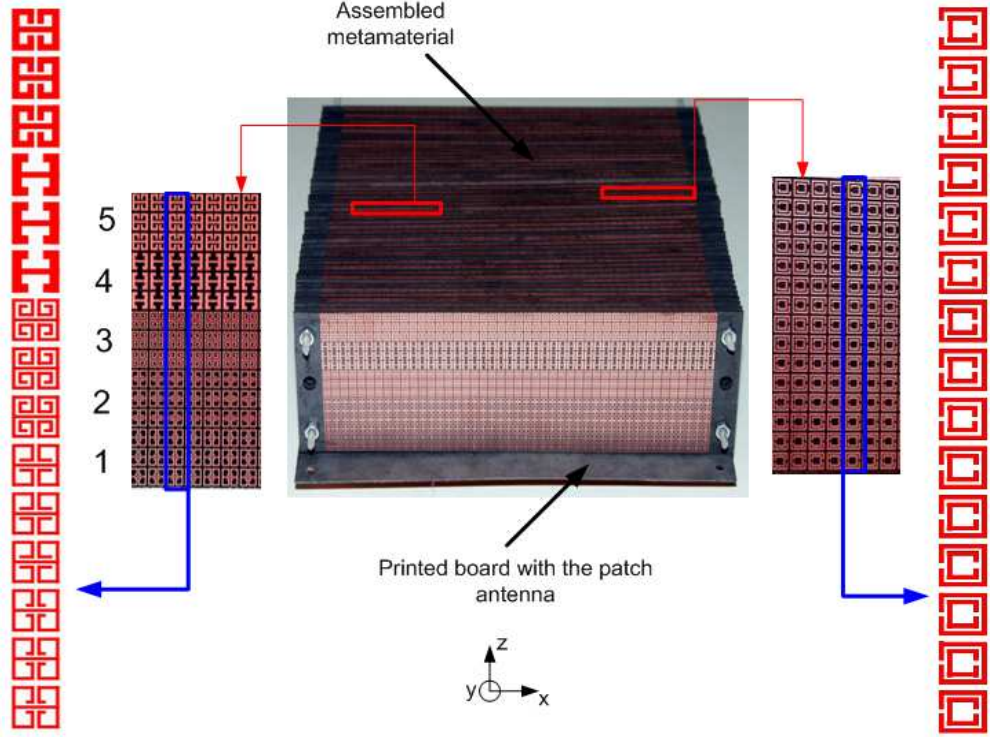

**Figure-3:** *Structure of the antenna: each magnetic and dielectric layer of the metamaterial is divided into 5 regions to assure the desired variations of electromagnetic parameters along wave propagation direction. The dimensions of the antenna are 15 cm x 15 cm x 5 cm. The operating frequency is 10.6GHz. Left and right inserts show details of the resonators used in the magnetic (right) and electric (left) metamaterial layers. Each level is made of three rows of identical resonators.*

The permeability ($\mu_{yy}$) and permittivity ($\varepsilon_{zz}$) parameter sets plotted in Fig. 2c can be respectively achieved in a composite metamaterial containing Pendry's split-ring resonators (SRRs)[47] and Smith's electric LC resonators (ELCs)[48], known to provide respectively a magnetic response and an electric response that can be tailored (Fig. 3). Because of constraints of the layout, we chose a rectangular unit cell with dimensions $p_x = p_z = 10/3$ mm for both resonators. The layout consisted of 5 regions, each of which was three unit cells long (10 mm). We were able to obtain the desired $\varepsilon_{zz}$ and $\mu_{yy}$ by tuning the resonators' geometric parameters as illustrated in Supplementary Figure S1. Using Ansoft's HFSS commercial full-wave finite-element simulation software, we performed a series of scattering (S) parameter simulations for the SRR and ELC unit cells separately over a discrete set of the geometric parameters covering the range of interest. A normal incident wave impinging on the unit cell is considered for simulations. Electric and magnetic symmetry planes are applied on the unit cell respectively for the faces normal to the electric and magnetic field vector. By calculating the unit cells separately, we therefore assume very low coupling between neighbouring ELCs and SRRs. The influence of this coupling is even lower when we consider mounting the electric and magnetic layers 2 by 2. A standard retrieval procedure[23] was then performed to obtain the effective material properties $\varepsilon_{zz}$ and $\mu_{yy}$ from the S-parameters. The discrete set of simulations and extractions was interpolated to obtain the particular values of the geometric parameters that yielded the desired material properties plotted in Fig. 1(c). Simulations were also realized under Comsol Multiphysics to assure the functionality of our metamaterial. We chose an operating frequency around 10 GHz, which yields a reasonable effective medium parameter $\lambda/p_x > 10$, where $\lambda$ is the free space wavelength.

In the designs presented in Fig. 2 and 3, we make use of SRRs and ELCs to realize the continuous-material properties required by the directive antenna. To illustrate the equivalence between continuous materials and actual combination of SRRs and ELCs metamaterials, we simulated the ideal antenna composed of continuous materials and the experimental antenna composed of SRRs and ELCs metamaterials simultaneously and we compared their electromagnetic properties. However, full-wave simulation of the experimental antenna is impossible using current computer resources owing to the extremely large memory and computing time required. Instead, full-wave simulations were done using the equivalent discrete material having parameters shown in Fig 2c. The full-wave simulations have been performed using finite-element method based commercial software Comsol Multiphysics. Also, the simulations have been made in a 2D configuration using RF module in a transverse electric wave propagation mode. A surface current having similar dimension as the patch feed is used to model the source. The diagram pattern of our antenna is plotted by inserting match boundaries with far field conditions. For the metamaterial, values of permittivity and permeability show in Fig. 2 have been introduced in each of the 5 layers. Figure 4 shows simulation results of the electric field emanating from the antenna in the continuous and discrete material cases. Excellent qualitative agreement is observed from the simulations, indicating that the SRRs-ELCs combination present nearly the same electromagnetic parameters as the continuous material. As it can observed the intensity of emitted radiation decreases rapidly since the source transformation operates only in the x-y plane.

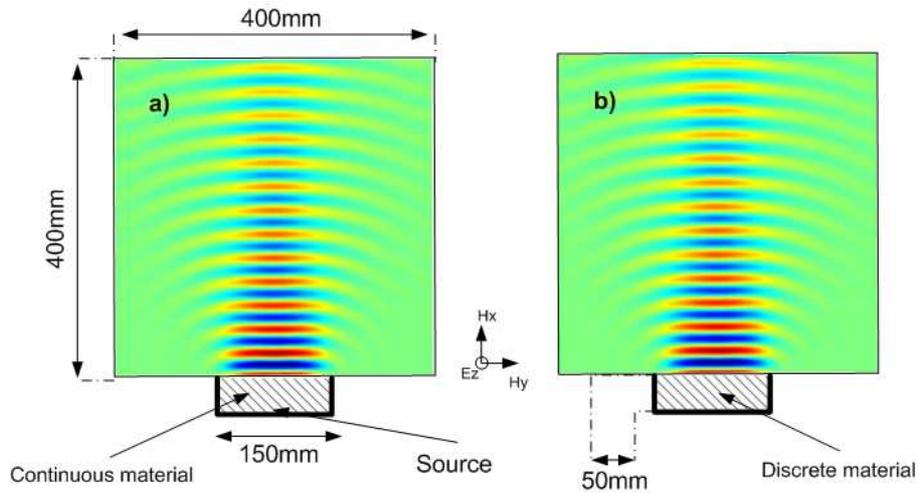

**Figure-4:** *Full-wave finite element simulations of electric fields emitted by the metamaterial antenna. Calculations are performed using the continuous and discrete materials in a 2D configuration using a line source as excitation. (a) Continuous material. (b) Discrete material.*

**Experimental setups and demonstration of ultra-directive emission**

To validate the directive emission device performances, two experiment systems are set up to measure the radiated field. The first method consists in measuring the far-field radiation patterns of the antenna in a fully anechoic chamber. Fig. 5a shows the far-field measurement system. In such emission-reception setup, the fabricated metamaterial antenna is used as emitter and a wideband (2-18 GHz) dual polarized horn antenna is used as the receiver to measure the radiated power level of the emitter. The measurements are done for computer-controlled elevation angle varying from -90° to +90°. The microwave source is a vector network analyser (Agilent 8722 ES) that we also use for detection. The feeding port is connected to the metamaterial antenna by means of a coaxial cable whereas the detecting port is connected to the horn antenna also by means of a coaxial cable. The measured far-field radiation pattern in the E-plane (plane containing E and k vectors) is presented in Fig. 5b.

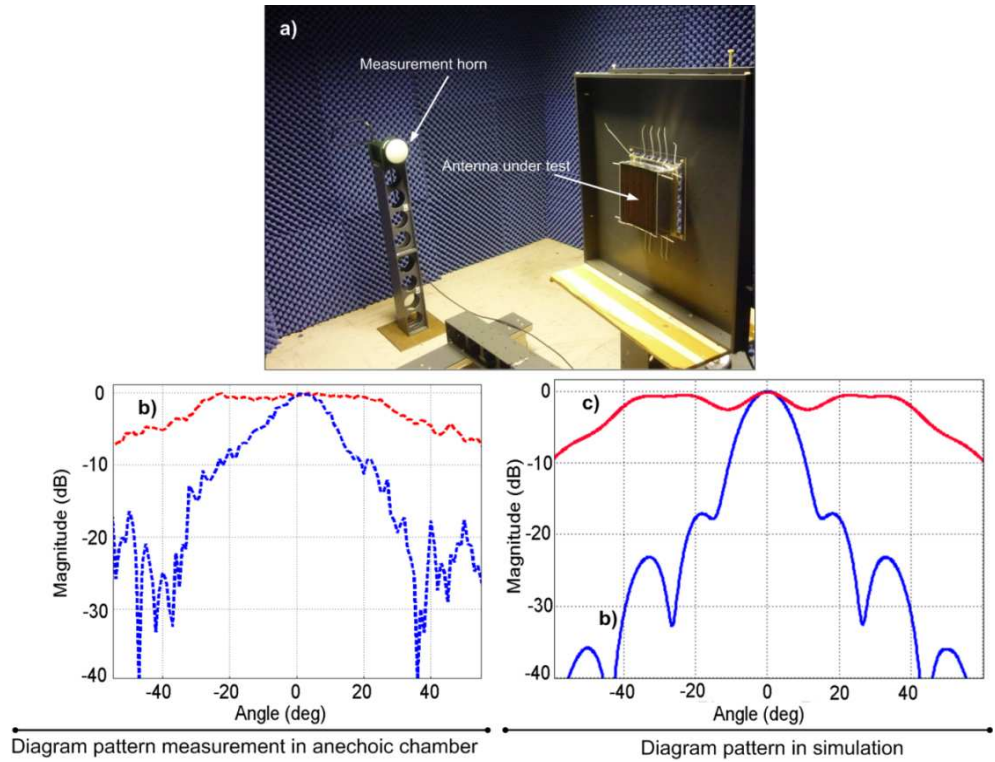

**Figure-5:** *Far-field measurement in a fully anechoic chamber. (a) Experimental setup system. (b) Measurements. (c) Simulations. Radiation patterns of the metamaterial antenna (blue trace) and of the feeding microstrip patch antenna alone (red trace) are presented at 10.6 GHz.*

The antenna presents maximum radiated power at 10.6 GHz with a directive main beam and low parasitic secondary lobes, under -15dB. The main lobe presents 13 degrees half-power beamwidth in the E-plane (x-y plane). This narrow beam width is less than that of a parabolic reflector antenna having similar dimensions (diameter equal to 15cm), where the half power beam width is around 16 degrees. Measurements are found to be consistent with the predicted radiation patterns shown in Fig. 5c.

The second experimental setup (Fig. 6(a)) was intended to scan the antenna's near field microwave radiation. The E-field is scanned by a field-sensing monopole probe connected to the network analyser by a coaxial cable. The probe was mounted on two orthogonal linear translation stages (computer-controlled Newport MM4006), so that the probe could be translated with respect to the radiation region of the antenna. By stepping the field sensor in small increments and recording the field amplitude and phase at every step, a full 2D spatial field map of the microwave near field pattern could be acquired in the free-space radiation region. The total scanning area covers $400 \times 400$ mm$^2$ with a step resolution of 2 mm in lateral dimensions shown by red arrows in Fig. 6a. Microwave absorbers are applied around the measurement stage in order to suppress undesired scattered radiations at the boundaries.

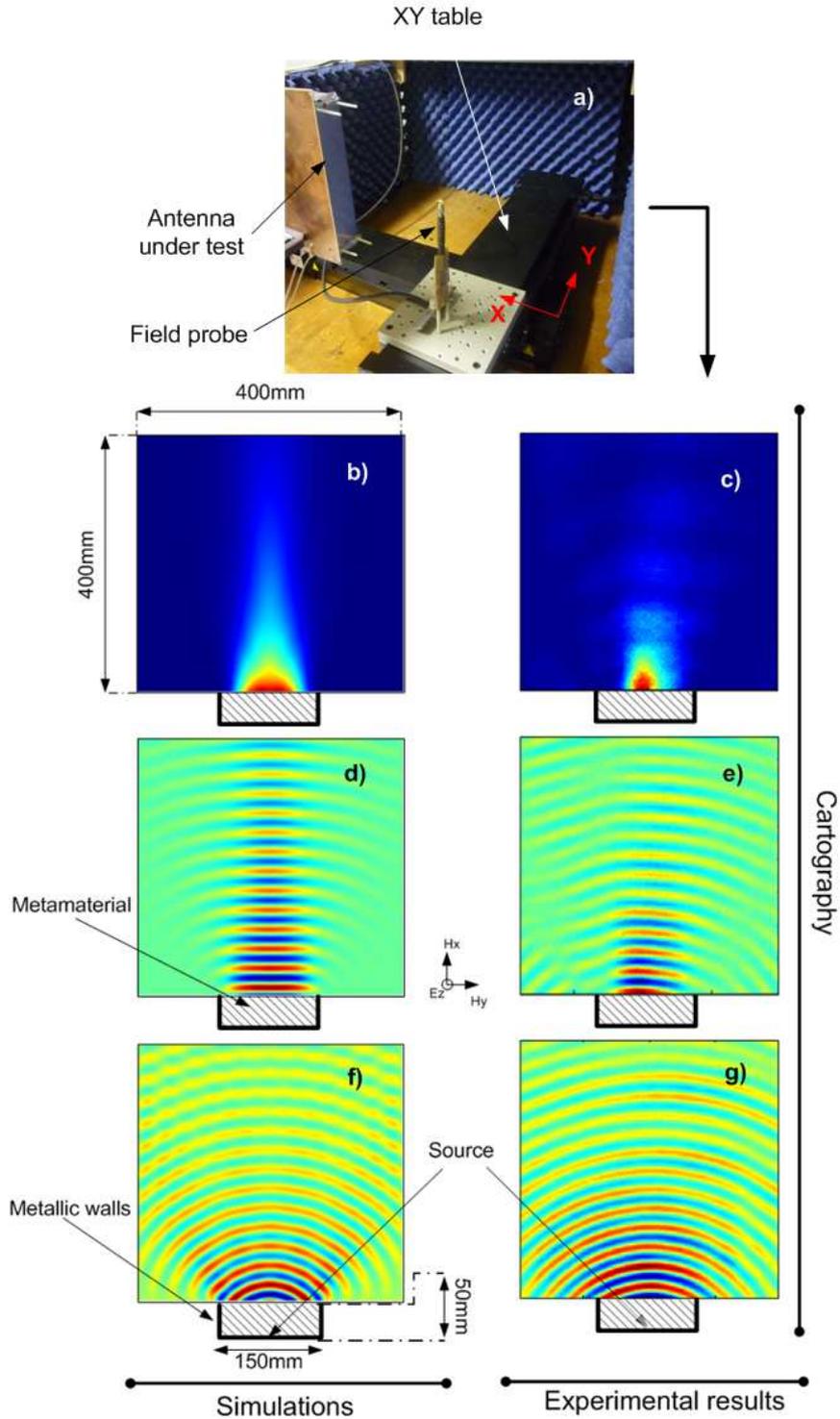

**Figure-6:** *Near-field scanning experiment in comparison with simulations (a) Experimental setup system. (b) Magnitude of the predicted Poynting vector (c) Magnitude of the experimental Poynting vector. (d) Magnitude of the predicted near field. (e) Mapping of the near-field. (f) Magnitude of the excitation source's predicted near field. (g) Mapping of the excitation source's near field. The mappings are shown at 10.6 GHz.*

Figure 6 shows the comparison between simulations and experimental results. In Fig. 6b, the magnitude of the numerical Poynting vector interpreted as an energy flux for the electromagnetic radiation is plotted for the device and compared to measurements in Fig. 6c. As stated earlier, the emission decreases rapidly since only the x-y plane has been considered for the source transformation procedure. A clear directive emission is radiated by the antenna as presented by the numerical simulation in Fig. 6d and measurement presented in Fig. 6e for the electric near field mapping of the antenna's radiation. Also, when compared to the radiation of the patch feed alone shown in Fig. 6f and 6g, we can note the narrow beam profile of our proposed device.

**Conclusion**

In summary, we designed, fabricated and measured a metamaterial-based ultra-directive emission by using the transformation optics approach in the microwave frequency regime. The device is engineered by transforming an isotropic source radiating in a cylindrical space to a directive one radiating in a rectangular space. It is composed of a feeding source covered by an anisotropic composite metamaterial cover. Both electric and magnetic parameters of this metamaterial are finely adjusted to correspond to the calculated parameters given by the transformation. Full-wave simulation results showed that the SRRs-ELCs metamaterials present nearly the same electromagnetic properties as the theoretical continuous materials. As a consequence, the artificial metamaterials used in the experimental device do have the bulk material behaviors that are expected. Calculations and measurements have shown a directive emission making this antenna competitive with conventional parabolic reflector ones. Synthesizing transformations optics through metamaterials appears to be promising and essential in the near future to control emission in a wide range of telecommunication applications.

**Acknowledgements**

The authors thank A. Degiron for fruitful discussions and his significant remarks on the manuscript.